\documentclass[aps,prl,twocolumn,groupedaddress,amsmath,amssymb]{revtex4-1}
\usepackage{fullpage}
\usepackage{mathtools}
\usepackage{algorithmic}
\usepackage{amsfonts}
\usepackage{amsmath}
\usepackage{amsthm}
\usepackage{amssymb}
\usepackage{graphics}
\usepackage{graphicx}
\usepackage{bm}
\usepackage{color}
\usepackage{hyperref}
\usepackage{fancyhdr}
\usepackage{footnote}
\usepackage{comment}
\usepackage[utf8]{inputenc}

\newtheorem{theorem}{Theorem}
\newtheorem{lemma}[theorem]{Lemma}
\newtheorem{definition}[theorem]{Definition}

\begin{document}

\title{Minimal physical resources for the realisation of measurement-based quantum computation}

\author{Monireh Houshmand$^{1,2}$}
\affiliation{Singapore University of Technology and Design, 8 Somapah Road, Singapore 487372}
\affiliation{Centre for Quantum Technologies, National University of Singapore, Block S15, 3 Science Drive 2, Singapore 117543}
\author{Mahboobeh Houshmand$^{1,2}$}
\email{Corresponding author: mahboobeh@sutd.edu.sg}
\affiliation{Singapore University of Technology and Design, 8 Somapah Road, Singapore 487372}
\affiliation{Centre for Quantum Technologies, National University of Singapore, Block S15, 3 Science Drive 2, Singapore 117543}
\author{Joseph F. Fitzsimons$^{1,2}$}
\affiliation{Singapore University of Technology and Design, 8 Somapah Road, Singapore 487372}
\affiliation{Centre for Quantum Technologies, National University of Singapore, Block S15, 3 Science Drive 2, Singapore 117543}
\begin{abstract}
In measurement-based quantum computation (MBQC), a special highly-entangled state (called a resource state) allows for universal quantum computation driven by single-qubit measurements and post-measurement corrections. Physical realisations of this model have been achieved in various physical systems for low numbers of qubits. The large number of qubits necessary to construct the resource state constitutes one of the main down sides to MBQC. However, in some instances it is possible to extend the resource state on the fly, meaning that not every qubit must be realised in the devices simultaneously. We consider the question of the minimal number of physical qubits that must be present in a system to directly implement a given measurement pattern. For measurement patterns with $n$ inputs, $n$ outputs and $m$ total qubits which have flow, we show that only $min (n+1,m)$ qubits are required, while the number of required qubits can be as high as $m-2$ for measurement patterns with only gflow. We discuss the implications of removing the Clifford part of a measurement pattern, using well-established transformation rules for Pauli measurements, for the presence of flow versus gflow, and hence the effect on the minimum number of physical qubits required to directly realise the measurement pattern.
\end{abstract}
\pacs{Valid PACS appear here}
\maketitle

The circuit model of quantum computation~\cite{deutsch1989quantum} provides a direct analogue to the common classical computational model based on networks of logic gates. On the other hand, measurement-based quantum computation (MBQC), first proposed by Raussendorf and Briegel in 2001~\cite{raussendorf2001one}, provides a conceptually and practically different model. This model harnesses unique features of quantum mechanics related to entanglement and measurement, and hence does not have a direct classical counterpart. In MBQC, computation is performed by making single qubit measurements on a special resource state, consisting of qubits prepared in a specific entangled state. As each measurement result is obtained, it is used to compute corrections to be taken into account in the bases of subsequent measurements. Due to the irreversible nature of the measurement process, the model is frequently referred to as one-way. Although two-dimensional cluster states were first suggested as the resource for universal quantum computation in the MBQC model~\cite{raussendorf2001one}, it was later shown that more general graph states could also be used~\cite{hein2004multiparty,danos2004robust}.

A measurement-based computation can be represented by a measurement pattern, which captures the structure of the resource state, the measurement angles assuming all non-output measurement results are zero, and a dependency structure used to adapt measurement angles based on non-zero measurement outcomes. The entanglement operations in a measurement pattern can be represented by a graph, where each vertex corresponds to a qubit and each edge corresponds to an entangling operation performed between the qubits indicated by the vertices it connects. This graph together with identified sets of input and output qubits is known as the \textit{open graph} corresponding to the computation \footnote{In the rest of the paper, qubits and vertices in open graphs will be used interchangeably.}. Since the measurements underlying such computations do not have predetermined outcomes, it is necessary to have some dependency structure in order to guarantee determinism. The existence of such a structure for arbitrary choices of measurement angles is determined fully by the open graph. For open graphs the presence of flow~\cite{danos2006determinism} is a sufficient condition, and generalized flow (gflow)~\cite{browne2007generalized} is a sufficient and necessary condition, for the existence of an appropriate dependency structure to ensure determinism. As gflow is more general than flow, in the rest of the paper we will use gflow only in reference to open graphs which do not have flow.

The unique features of MBQC have made it a natural choice in many quantum computer architectures~\cite{nielsen2004optical,browne2005resource,benjamin2006brokered,friesen2008one,blythe2006cavity}. It has also emerged as a useful theoretical tool, particularly in relation to the design of secure computing protocols as blind quantum computing (BQC)~\cite{broadbent2009,fitzsimons2012unconditionally,morimae2013blind}. Several other tasks such as entanglement purification in the presence of noise and imperfections~\cite{zwerger2013universal} and quantum error correction (QEC)~\cite{zwerger2014hybrid} can be achieved very efficiently in a measurement-based way, i.e., with resource states of minimal size. Furthermore, MBQC allows for topological fault-tolerance to be realised in a very direct and beautiful way~\cite{raussendorf2007topological}.

However, despite the advantages of the MBQC model, its realisation is often expensive in terms of physical qubits, as the number of qubits in a measurement pattern is usually much more than the number of logical qubits in the computation. This stems from the fact one qubit is required for each (non-Clifford) single qubit gate in the computation. For example, an instance of the three-qubit quantum Fourier transform (QFT), considered in Ref.~\cite{hein2004multiparty}, has a realisation requiring 33 qubits in a graph state. Moreover, some applications, such as verification in blind quantum computation protocols~\cite{fitzsimons2012unconditionally,kashefi2015optimised,morimae2016measurement}, may significantly increase the number of required qubits. MBQC has been demonstrated experimentally using various discrete-variable (qubit) systems~\cite{walther2005experimental,chen2007experimental,prevedel2007high,vallone2008active,tokunaga2008generation,yao2012experimental,lanyon2013measurement} and continuous variable systems~\cite{miwa2009demonstration,ukai2011demonstration,pooser2014continuous}. However, experiments for qubit systems have generally been restricted to low numbers of qubits and scaling them up is an important challenge~\cite{walther2005experimental,pooser2014continuous}.

Here we examine the number of physical qubits required to realise a measurement pattern, when entanglement operations and measurements can be re-ordered. We consider the question of whether the whole resource state has to be constructed at the beginning, or whether it is possible to add qubits on an as needed basis. In the latter case, we consider the minimal number of necessary physical qubits at any time, which we denote $min_{QR}$. We show that $min_{QR}$ is different for open graphs with flow versus those with only gflow, and in some instances this difference can be dramatic. There is a well established method for translating from quantum circuits to measurement patterns through the use of gate teleportation~\cite{childs2005unified}. Our results can be thought of as providing a sensible method for doing the reverse translation, from measurement pattern to circuit model, even for patterns which may have been created without reference to circuits. This is particularly in the case of blind and verifiable quantum computing protocols naturally constructed in the measurement based model~\cite{broadbent2009,fitzsimons2012unconditionally,morimae2013blind}, providing a way to implement such protocols in devices which directly implement the circuit model with far lower qubit resources. It also provides a mechanism to take advantage of phenomena such as flow ambiguity~\cite{mantri2016} directly in the circuit model augmented with individual gate teleportations.

The remainder of the paper is structured as follows. We begin by introducing needed definitions and background. We then derive the required physical qubit resources for measurement-based computations for the cases of flow and gflow. We conclude with the examination of the effect of removing Pauli measurements, which implement Clifford group gates, in terms of its effect on the presence of flow.

Following the notation of~\cite{danos2007measurement}, a measurement-based computation can be represented by a measurement pattern, or simply pattern. A pattern is defined as \emph{P = (V, I, O, A)}, where \emph{V} is the set of qubits, \emph{I} $ \subseteq $ \emph{V} and \emph{O} $ \subseteq $ \emph{V} are two possibly overlapping sets representing the inputs and outputs of the computation respectively and \emph{A} is a finite set of operations which act on \emph{V} as defined in the following:
\begin{itemize}
\item 1-qubit auxiliary \emph{preparation} $Pr_v$ prepares a qubit \emph{v} $ \in $ \emph{V} in the state $\frac{1}{\sqrt{2}}\left(\left\vert 0 \right\rangle+\left\vert 1 \right\rangle\right)$,
\item 2-qubit \emph{entanglement operation} $E_{uv}$ performs a CZ operation on qubits \emph{u,v} $ \in$ \emph{V},
\item 1-qubit \emph{correction operations}  $X_v$ and $Z_v$ apply Pauli \emph{X} and \emph{Z} corrections on qubit \emph{v}, and
\item 1-qubit \emph{measurement operation} $M_v^\alpha$ measures the qubit \emph{v} in the orthonormal basis of $\left| { \pm \alpha } \right\rangle : = \frac{1}{{\sqrt 2 }}\left(\left| 0 \right\rangle  \pm e^{i\alpha } \left| 1 \right\rangle \right)$, where $\alpha \in [0,2\pi)$ is called the angle of measurement.
\end{itemize}

For a graph $G=(V, E)$, $V$ denotes the set of its vertices and $E$ is the set of its edges. An open graph is a triplet $(G,I,O)$, where $G=(V,E)$ is an undirected graph and $I, O \subseteq V$ are respectively the sets of input and output vertices. The \textit{size} of $G$ is its number of vertices is denoted by $m$. Non-input vertices are denoted by $I^C$ (the complement of $I$ in the graph) and non-output vertices are denoted by $O^C$ (the complement of $O$ in the graph).

Flow and gflow on open graphs, as defined in the following, determine an ordering of measurements which guarantees that measurement angles can always be adapted based on previous results to implement a unitary transformation deterministically, for any choice of measurement angles.

\begin{definition}[Danos \& Kashefi~\cite{danos2006determinism}]
An open graph $(G,I,O)$ has \emph{flow} if and only if there exists a map $f: O^C \to I^C$ and a strict partial order $\prec_f$ over $V$ such that all of the following conditions hold for all $i \in O^C$.
\begin{itemize}
  \item $i \prec_f f(i)$,
  \item if $j \in N(f(i))$, then $j=i$ or $i \prec_f j$, where $N(v)$ contains adjacent vertices of $v$ in $G$,
  \item $i \in N(f(i))$.
\end{itemize}
In this case, ($f, \prec_f$) is called a flow on $(G,I,O)$.
\end{definition}

To aid clarity, we will make use of the notation $u \to v$, if $f(u)=v$ and $u \Rightarrow v$, if $u \rightarrow v_1 \rightarrow v_2\,\rightarrow \ldots \rightarrow\, v_{n-1} \rightarrow v_n$ where $v_n=v$.

\begin{definition}[Browne et al.~\cite{browne2007generalized}]
An open graph $(G,I,O)$ has \emph{generalised flow (gflow)} if and only if there exists a map $g: O^C \to P^{I^C }$ (the set of all subsets of vertices in $I^C$) and a strict partial order $\prec_g$ over $V$ such that all of the following conditions hold for all $i \in O^C$.
\begin{itemize}
  \item if $j \in g(i)$ then $i \prec_g j$,
  \item if $j \in \emph{Odd}(g(i))$, then $j=i$ or $i \prec_g j$, where $\emph{Odd}(K) = \{ k|\, \left| {N (k) \cap K} \right| = 1\,\,\bmod \,\,2\}$,
  \item $i \in \emph{Odd}(g(i))$.
\end{itemize}
In this case, $(g, \prec_g)$ is called a gflow on $(G,I,O)$.
\end{definition}

Let $(G, I, O)$ be an open graph with flow. Then, a structure called \emph{path cover}~\cite{de2008finding} is induced in $G$ as defined in the following. A collection $P_f$ of directed paths in $G$ is called a path cover of $(G,I,O)$ if (i) each $v \in V$ is included in exactly one path, in other words paths are vertex-disjoint and they cover $G$, (ii) each path in $P_f$ is either disjoint from $I$ or intersects $I$ only at its initial vertex, and (iii) each path in $P_f$ intersects $O$ only at its final vertex. In this paper, we assume that $\left| I \right| = \left| O \right|=n$ (corresponding to patterns performing unitary transformations). In this case, for $(G,I,O)$, there are $n$ paths, each starting from an input vertex, $i_j$, and ending at an output vertex, $o_j$ (possibly overlapping), such that $i_j \rightarrow v_{1j} \rightarrow  v_{2j}\, \rightarrow \ldots \rightarrow\, v_{n_{jj}} \rightarrow  o_j \in P_f$. The path to which qubit $w$ belongs is denoted by $\mathcal{P}(w)$.

Now, we consider the reordering of the entanglement and measurement operations such that the number of physical qubits necessary at any one time is minimised. The idea is based on postponing each entangling operation as long as possible. Suppose it is the turn of a qubit $w \in O^C$ to be measured with respect to an ordering of measurements induced by flow. We will denote the set of unmeasured qubits at this stage, excluding $w$, as $\mathcal{U}_w$ and the set of measured qubits as $M_w$. The measurement on a particular qubit, $w$ commutes with entangling operations between $u$ and $v$ when neither $u$ nor $v$ is equal to $w$, but does not commute with entanglement operations between $w$ and its unmeasured neighbours~\cite{nikahd2015one}. Therefore, these operations have to be performed first before the measurement. The set of unmeasured neighbours of $w$ is denoted by $\mathcal{N}_w$, which is equal to $N(w) \cap \mathcal{U}_w$. The measurement of the qubit $w$ affects the state of qubits in $\mathcal{N}_w$. As no operation acts on a previously measured qubit~\cite{danos2007measurement}, $w$ is not required beyond this point during the realisation of a pattern.

Now, we investigate the minimal set of qubits which must simultaneously exist prior to the measurement of $w$, excluding $w$ itself, which we label $\mathcal{Q}_w$. This set is the union of two subsets of vertices: (i) the subset that is required for performing the measurement on $w$, $\mathcal{N}_w$, and (ii) the subset of qubits which have been affected by previous operations and which have not been measured, and hence must be retained until measurement (if they do not belong to $O$) or until the end of computation. We now characterise this latter subset.

At the beginning of a measurement-based computation, the qubits in $I$ are provided or prepared in some joint input state and must be retained until they are measured (if they do not belong to O), or until the end of computation. When it is the turn of a qubit $w$ to be measured, the set of all unmeasured input qubits excluding $w$ is denoted $\mathcal{I}_w$. During the computation, measurements cannot be commuted past entangling operations involving the same qubit, and hence the neighbours of any measured qubits must either be measured or retained. We will denote by $\mathcal{O}_w$ the subset of qubits in $\mathcal{U}_w$ with measured neighbours. More formally, $\mathcal{O}_w=\{v\in\mathcal{U}_w |  N(w) \cap  \mathcal{M}_w \neq \varnothing \}$, where $\varnothing$ is the empty set. Therefore, we have $\mathcal{Q}_w=\mathcal{N}_w \cup \mathcal{I}_w \cup \mathcal{O}_w$.

Suppose it is the turn of a qubit $w \in O^C$ to be measured with respect to an ordering of measurements induced by flow. Then, the following statement holds.

\begin{lemma}
\label{nahayee}
Let $(G,I,O)$ be an open graph with flow. There exists exactly one member of $\mathcal{Q}_w$ in each path $\mathcal{P}$ of $P_f$.

\begin{proof}
We first prove that in each $\mathcal{P}$ there exists at least one member of $\mathcal{Q}_w$, and then we prove that this lower bound must be saturated. We will use $v$ to label this unique vertex for a particular path.

Tackling the upper bound first, for a given $\mathcal{P}$, one of the following two cases will happen:
\begin{enumerate}
\item $w \in \mathcal{P}$:
With respect to the flow definition, there is $v \in \mathcal{N}_w\cap \mathcal{U}_w$ given by $v=f(w)$ such that $\mathcal{P}(v)=\mathcal{P}(w)$.
\item $w \notin \mathcal{P}$: In this situation, there are only two possible cases:
 \begin{itemize}
 \item None of the qubits in $\mathcal{P}$ have been measured previously. Therefore, there exists $v \in \mathcal{I}_w$ in this path.
 \item At least one of the qubits in $\mathcal{P}$ has been measured previously. Let $u$ be the last qubit which has been measured in this path. Therefore, we have $v=f(u) \in \mathcal{O}_w$.
 \end{itemize}
\end{enumerate}
This guarantees that at least one qubit in each path must be in $\mathcal{Q}_w$, when the input state is left unspecified.

We now show that if  $u, v \in \mathcal{Q}_w$, and $u \ne v$, then $\mathcal{P}(u) \ne \mathcal{P}(v)$. The proof is done by contradiction. Suppose $\mathcal{P}(u)=\mathcal{P}(v)$ and without loss of generality, suppose $u \Rightarrow v$. In such a situation, it must be the case that $v \notin \mathcal{I}_w$. Therefore, one of the following two cases will occur:
\begin{enumerate}
\item $v \in \mathcal{N}_w$: Based on the flow definition, $u$ has to be measured before $w$ which belongs to $N(v)$. Therefore, $u \notin \mathcal{Q}_w$.
\item $v \in \mathcal{O}_w$: Based on the flow definition, $u$ has to be measured before all of the neighbours of $v$, but since $v \in \mathcal{O}_w$, a neighbour of $v$ has been previously measured. Therefore, $u \notin \mathcal{Q}_w$.
\end{enumerate}
This leads directly to the conclusion that in each $\mathcal{P}$, $v$ is the unique member of $\mathcal{Q}_w$.
\end{proof}
\end{lemma}

In Theorem~\ref{theo1}, $min_{QR}$ is determined for open graphs with flow.

\begin{theorem}\label{theo1} Let $(G,I,O)$ be an open graph with flow, with the same number of inputs and outputs, $n$. To realise patterns with the underlying open graph, $min_{QR}$ is $\min(n+1,m)$, where $m$ is the whole number of qubits in the pattern.
\begin{proof}
First, consider the case that $I=O\,(m=n)$. In this case, $min_{QR}$ is trivially equal to $m=n$. Now, suppose that $I \ne O$, and in this case, according to Lemma~\ref{nahayee}, the size of $\mathcal{Q}_w$ is equal to the number of paths in the graph, trivially equal to $n$, and therefore by including the presence of $w$, we have $min_{QR}= n+1$.\end{proof}
\end{theorem}

\begin{figure}
  \centering
    \includegraphics[width=0.21\textwidth]{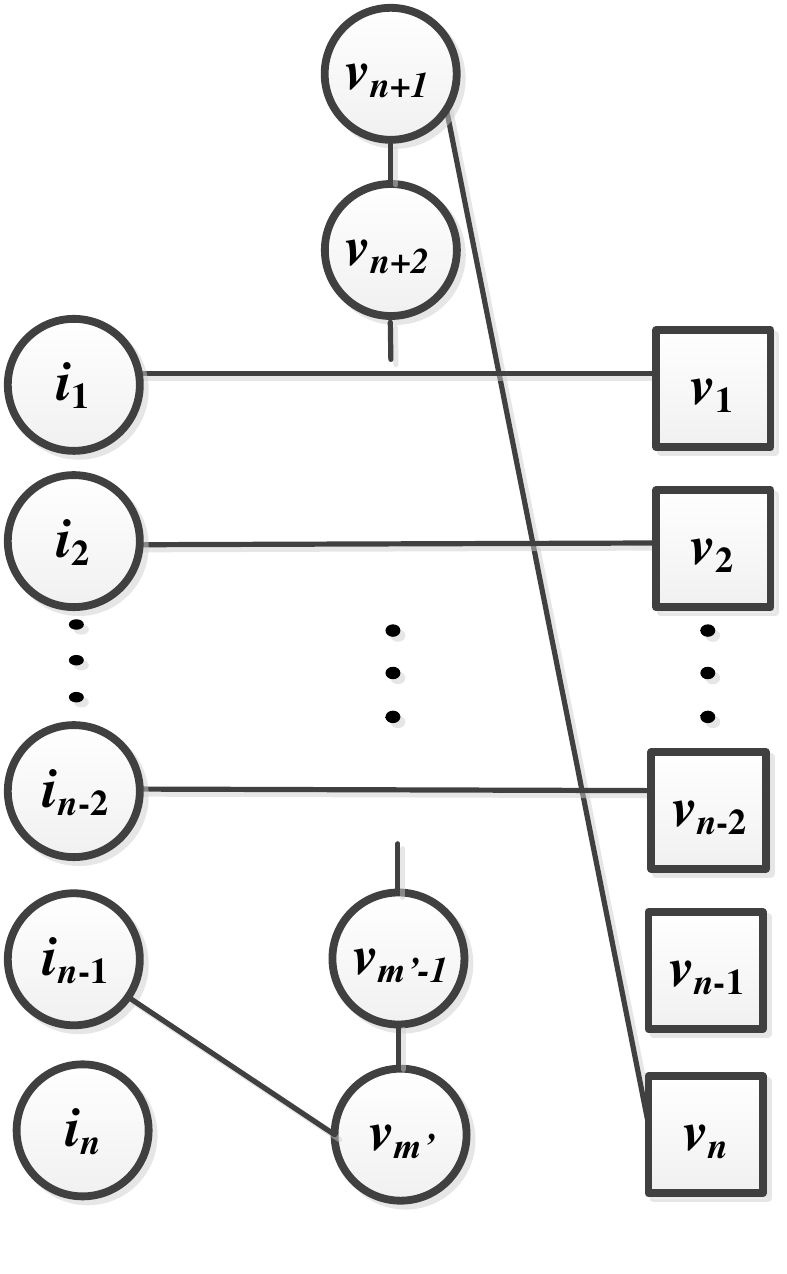}
  \caption{Representation of $(H^C_n,I,O)$. Input qubits are shown by $i_1,i_2,...,i_n$ and squared vertices represent output qubits. }
  \label{fig:Fig1}
\end{figure}

Although we have shown that $min_{QR}$ for open graphs with flow on $n$ inputs is $min(n+1,m)$, it is not the case for open graphs with gflow.
This is demonstrated by constructing a family of open graphs which require large numbers of qubits to be present as a counter-example.
We will consider open graphs $(H_n,I,O)$ with $n>1$ inputs, $\{i_1,i_2,...,i_n\}$, $n$ outputs, $\{v_1,v_2,...,v_n\}$, and $(m-2n) \ne 0$ intermediate qubits, $\{v_{n+1},v_{n+2},...,v_{m'}\}$, where $m'=m-n$. Rather than specifying the edges of $H_n$ directly, we instead specify the edges of the graph $H^C_n$ obtained by edge complement of $H_n$. This is for simplicity since $H_n$ will be highly connected. The graph $H^C_n$, shown in Fig.~\ref{fig:Fig1}, has the following edges:
$\{i_j ,v_j \}$ for $j \in \{ 1,2,...,n - 2\}$, $\{v_{n+j},v_{n+j+1}\}$ for $j \in \{0, 1, ..., m'-n-1\}$, and $\{i_{n-1},v_{m'}\}$.

A gflow on $H_n$ can be found by applying the algorithm proposed in Ref.~\cite{mhalla2008finding}, which yields the following: $g(i_j)=\{v_j,v_{n-1}\}$ for $j\in \{1,...,n-2\}$, $g(v_{j})=\{v_{j-2},v_{j-1}\}$ for $j \in \{n+1,...,m'\}$, $\,g(i_{n-1})=\{v_{m'-1},v_{m'}\}$ and $g(i_n)=v_{m'}$. Since from Fig.~\ref{fig:Fig1} the maximum degree of $H^C_n$ can easily be seen to be $2$, the minimal degree of $H_n$ must be equal to $m-3$. Starting from a qubit $w$ in a partial order induced by a gflow on this open graph, we have $\left| {\mathcal{N}_w } \right| \ge m - 3$. Therefore $min_{QR} \ge m-2$.

We conclude by examining the effect of measurement of Pauli operators on graphs with flow and those with gflow, since this can alter the presence of flow. Unitary operators which map Pauli group operators to the Pauli group under conjugation are known as Clifford group operations. Any of these operators can be implemented by patterns with Pauli measurements $X$ and $Y$ only~\cite{browne2006one}. Due to the nature of corrections made during an MBQC, measurements of Pauli operators are unaffected and can be shifted to the start of the computation. In Ref.~\cite{hein2004multiparty}, general transformation rules for graphs are described when Pauli measurements are performed on qubits. This allows for Pauli measurements to be eliminated by modifying the graph state to be prepared and updating the other measurement bases. For example, in the case of a $Y$ measurement on qubit $w$, the graph corresponding to the resulting state is obtained by replacing the subgraph consisting of neighbours of $w$ by its complement, and removing $w$ and any incident edges. Measurement bases of qubits neighbouring $w$ also need to be updated.

\begin{figure}
  \centering
    \includegraphics[width=0.48\textwidth]{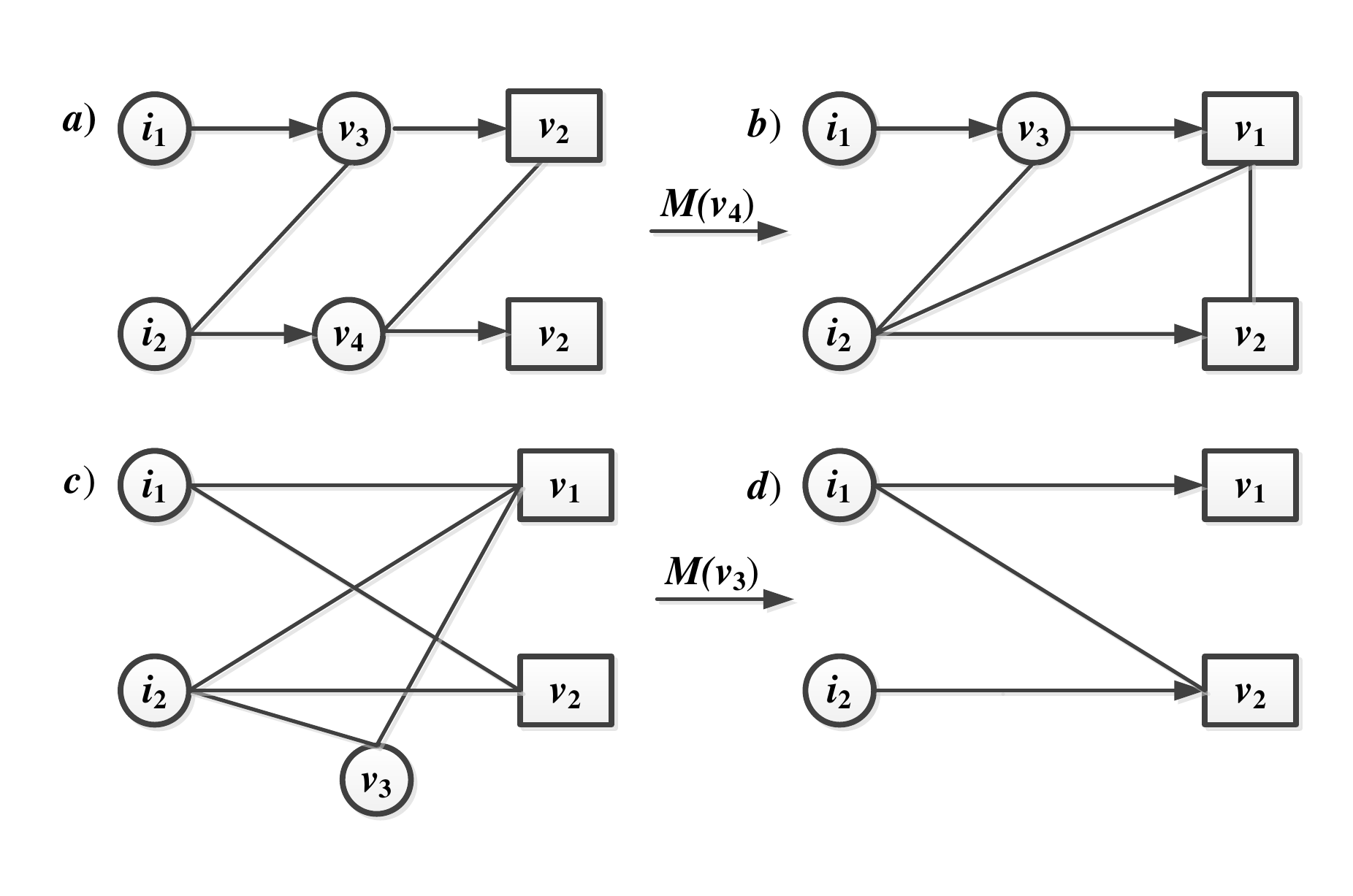}
  \caption{Examples of removing or introducing flow in open graphs after measuring a single qubit in the $Y$ basis. Input qubits are shown by $i_1,i_2$ and squared vertices represent output qubits. a) A sample open graph, $(G_a,I,O)$ with flow. b) The resulting open graph after measuring $v_4$ in $(G_a,I,O)$, which has flow. c) A sample open graph $(G_c,I,O)$ with gflow. d) The resulting open graph with flow after measuring $v_3$ in $(G_c,I,O)$.}
  \label{fig:Fig2}
\end{figure}

Consider an open graph $(H'_n,I,O)$ where $H'_n$ is a graph consisting of $H^C_n$ (shown in Fig.~\ref{fig:Fig1}) and another vertex, $y$ which is connected to all of the vertices of $H_n^C$. $(H'_n,I,O)$ has a flow as follows: $f(i_j)=v_j$ for $j \in \{1,...,n-2\}$, $f(i_{n-1})=v_{m'}$, $f(i_n)=y$, $f(v_{j})=v_{j-1}$ for $j \in \{n+1,...,m'\}$, and $f(y)=v_{n-1}$. Thus, $min_{QR}=n+2$. It can be readily verified that when $y$ is measured in the $Y$-basis, $H_n'$ will be transformed to $H_n$, which has been previously shown that has gflow, with $min_{QR} \ge m-2$. On the other hand, when any vertex in $H_n$ is measured in the $Y$-basis, $(H_n,I,O)$ will lead to an open graph which has gflow but not flow. In Fig.~\ref{fig:Fig2}, further examples are given where measurement maintains flow and where Pauli measurement introduces flow to an open graph that previously had only gflow. This highlights the fact that when certain measurements are fixed to a Pauli basis in measurement pattern, their removal can have either a positive or negative effect on the minimal physical qubit resources necessary to implement the pattern.

\textit{Acknowledgements --} The authors thank Tommaso Demarie, Yingkai Ouyang, and Atul Mantri for useful comments on an earlier version of this paper. The second author is grateful to Eesa Nikahd for helpful discussions. The authors acknowledge support from Singapore's Ministry of Education and National Research Foundation, and the Air Force Office of Scientific Research under AOARD grant FA2386-15-1-4082. This material is based on research funded in part by the Singapore National Research Foundation under NRF Award NRF-NRFF2013-01.

\bibliography{apssamp}

\begin{thebibliography}{37}%
\makeatletter
\providecommand \@ifxundefined [1]{%
 \@ifx{#1\undefined}
}%
\providecommand \@ifnum [1]{%
 \ifnum #1\expandafter \@firstoftwo
 \else \expandafter \@secondoftwo
 \fi
}%
\providecommand \@ifx [1]{%
 \ifx #1\expandafter \@firstoftwo
 \else \expandafter \@secondoftwo
 \fi
}%
\providecommand \natexlab [1]{#1}%
\providecommand \enquote  [1]{``#1''}%
\providecommand \bibnamefont  [1]{#1}%
\providecommand \bibfnamefont [1]{#1}%
\providecommand \citenamefont [1]{#1}%
\providecommand \href@noop [0]{\@secondoftwo}%
\providecommand \href [0]{\begingroup \@sanitize@url \@href}%
\providecommand \@href[1]{\@@startlink{#1}\@@href}%
\providecommand \@@href[1]{\endgroup#1\@@endlink}%
\providecommand \@sanitize@url [0]{\catcode `\\12\catcode `\$12\catcode
  `\&12\catcode `\#12\catcode `\^12\catcode `\_12\catcode `\%12\relax}%
\providecommand \@@startlink[1]{}%
\providecommand \@@endlink[0]{}%
\providecommand \url  [0]{\begingroup\@sanitize@url \@url }%
\providecommand \@url [1]{\endgroup\@href {#1}{\urlprefix }}%
\providecommand \urlprefix  [0]{URL }%
\providecommand \Eprint [0]{\href }%
\providecommand \doibase [0]{http://dx.doi.org/}%
\providecommand \selectlanguage [0]{\@gobble}%
\providecommand \bibinfo  [0]{\@secondoftwo}%
\providecommand \bibfield  [0]{\@secondoftwo}%
\providecommand \translation [1]{[#1]}%
\providecommand \BibitemOpen [0]{}%
\providecommand \bibitemStop [0]{}%
\providecommand \bibitemNoStop [0]{.\EOS\space}%
\providecommand \EOS [0]{\spacefactor3000\relax}%
\providecommand \BibitemShut  [1]{\csname bibitem#1\endcsname}%
\let\auto@bib@innerbib\@empty
\bibitem [{\citenamefont {Deutsch}(1989)}]{deutsch1989quantum}%
  \BibitemOpen
  \bibfield  {author} {\bibinfo {author} {\bibfnamefont {D.}~\bibnamefont
  {Deutsch}},\ }in\ \href@noop {} {\emph {\bibinfo {booktitle} {Proceedings of
  the Royal Society of London A: Mathematical, Physical and Engineering
  Sciences}}},\ Vol.\ \bibinfo {volume} {425}\ (\bibinfo {organization} {The
  Royal Society},\ \bibinfo {year} {1989})\ pp.\ \bibinfo {pages}
  {73--90}\BibitemShut {NoStop}%
\bibitem [{\citenamefont {Raussendorf}\ and\ \citenamefont
  {Briegel}(2001)}]{raussendorf2001one}%
  \BibitemOpen
  \bibfield  {author} {\bibinfo {author} {\bibfnamefont {R.}~\bibnamefont
  {Raussendorf}}\ and\ \bibinfo {author} {\bibfnamefont {H.~J.}\ \bibnamefont
  {Briegel}},\ }\href@noop {} {\bibfield  {journal} {\bibinfo  {journal}
  {Physical Review Letters}\ }\textbf {\bibinfo {volume} {86}},\ \bibinfo
  {pages} {5188} (\bibinfo {year} {2001})}\BibitemShut {NoStop}%
\bibitem [{\citenamefont {Hein}\ \emph {et~al.}(2004)\citenamefont {Hein},
  \citenamefont {Eisert},\ and\ \citenamefont {Briegel}}]{hein2004multiparty}%
  \BibitemOpen
  \bibfield  {author} {\bibinfo {author} {\bibfnamefont {M.}~\bibnamefont
  {Hein}}, \bibinfo {author} {\bibfnamefont {J.}~\bibnamefont {Eisert}}, \ and\
  \bibinfo {author} {\bibfnamefont {H.~J.}\ \bibnamefont {Briegel}},\
  }\href@noop {} {\bibfield  {journal} {\bibinfo  {journal} {Physical Review
  A}\ }\textbf {\bibinfo {volume} {69}},\ \bibinfo {pages} {062311} (\bibinfo
  {year} {2004})}\BibitemShut {NoStop}%
\bibitem [{\citenamefont {Danos}\ \emph {et~al.}(2005)\citenamefont {Danos},
  \citenamefont {Kashefi},\ and\ \citenamefont {Panangaden}}]{danos2004robust}%
  \BibitemOpen
  \bibfield  {author} {\bibinfo {author} {\bibfnamefont {V.}~\bibnamefont
  {Danos}}, \bibinfo {author} {\bibfnamefont {E.}~\bibnamefont {Kashefi}}, \
  and\ \bibinfo {author} {\bibfnamefont {P.}~\bibnamefont {Panangaden}},\
  }\href@noop {} {\bibfield  {journal} {\bibinfo  {journal} {Physical Review
  A}\ }\textbf {\bibinfo {volume} {72}} (\bibinfo {year} {2005})}\BibitemShut
  {NoStop}%
\bibitem [{Note1()}]{Note1}%
  \BibitemOpen
  \bibinfo {note} {In the rest of the paper, qubits and vertices in open graphs
  will be used interchangeably.}\BibitemShut {Stop}%
\bibitem [{\citenamefont {Danos}\ and\ \citenamefont
  {Kashefi}(2006)}]{danos2006determinism}%
  \BibitemOpen
  \bibfield  {author} {\bibinfo {author} {\bibfnamefont {V.}~\bibnamefont
  {Danos}}\ and\ \bibinfo {author} {\bibfnamefont {E.}~\bibnamefont
  {Kashefi}},\ }\href@noop {} {\bibfield  {journal} {\bibinfo  {journal}
  {Physical Review A}\ }\textbf {\bibinfo {volume} {74}},\ \bibinfo {pages}
  {052310} (\bibinfo {year} {2006})}\BibitemShut {NoStop}%
\bibitem [{\citenamefont {Browne}\ \emph {et~al.}(2007)\citenamefont {Browne},
  \citenamefont {Kashefi}, \citenamefont {Mhalla},\ and\ \citenamefont
  {Perdrix}}]{browne2007generalized}%
  \BibitemOpen
  \bibfield  {author} {\bibinfo {author} {\bibfnamefont {D.~E.}\ \bibnamefont
  {Browne}}, \bibinfo {author} {\bibfnamefont {E.}~\bibnamefont {Kashefi}},
  \bibinfo {author} {\bibfnamefont {M.}~\bibnamefont {Mhalla}}, \ and\ \bibinfo
  {author} {\bibfnamefont {S.}~\bibnamefont {Perdrix}},\ }\href@noop {}
  {\bibfield  {journal} {\bibinfo  {journal} {New Journal of Physics}\ }\textbf
  {\bibinfo {volume} {9}},\ \bibinfo {pages} {250} (\bibinfo {year}
  {2007})}\BibitemShut {NoStop}%
\bibitem [{\citenamefont {Nielsen}(2004)}]{nielsen2004optical}%
  \BibitemOpen
  \bibfield  {author} {\bibinfo {author} {\bibfnamefont {M.~A.}\ \bibnamefont
  {Nielsen}},\ }\href@noop {} {\bibfield  {journal} {\bibinfo  {journal}
  {Physical review letters}\ }\textbf {\bibinfo {volume} {93}},\ \bibinfo
  {pages} {040503} (\bibinfo {year} {2004})}\BibitemShut {NoStop}%
\bibitem [{\citenamefont {Browne}\ and\ \citenamefont
  {Rudolph}(2005)}]{browne2005resource}%
  \BibitemOpen
  \bibfield  {author} {\bibinfo {author} {\bibfnamefont {D.~E.}\ \bibnamefont
  {Browne}}\ and\ \bibinfo {author} {\bibfnamefont {T.}~\bibnamefont
  {Rudolph}},\ }\href@noop {} {\bibfield  {journal} {\bibinfo  {journal}
  {Physical Review Letters}\ }\textbf {\bibinfo {volume} {95}},\ \bibinfo
  {pages} {010501} (\bibinfo {year} {2005})}\BibitemShut {NoStop}%
\bibitem [{\citenamefont {Benjamin}\ \emph {et~al.}(2006)\citenamefont
  {Benjamin}, \citenamefont {Browne}, \citenamefont {Fitzsimons},\ and\
  \citenamefont {Morton}}]{benjamin2006brokered}%
  \BibitemOpen
  \bibfield  {author} {\bibinfo {author} {\bibfnamefont {S.~C.}\ \bibnamefont
  {Benjamin}}, \bibinfo {author} {\bibfnamefont {D.~E.}\ \bibnamefont
  {Browne}}, \bibinfo {author} {\bibfnamefont {J.}~\bibnamefont {Fitzsimons}},
  \ and\ \bibinfo {author} {\bibfnamefont {J.~J.}\ \bibnamefont {Morton}},\
  }\href@noop {} {\bibfield  {journal} {\bibinfo  {journal} {New Journal of
  Physics}\ }\textbf {\bibinfo {volume} {8}},\ \bibinfo {pages} {141} (\bibinfo
  {year} {2006})}\BibitemShut {NoStop}%
\bibitem [{\citenamefont {Friesen}\ and\ \citenamefont
  {Feder}(2008)}]{friesen2008one}%
  \BibitemOpen
  \bibfield  {author} {\bibinfo {author} {\bibfnamefont {T.~P.}\ \bibnamefont
  {Friesen}}\ and\ \bibinfo {author} {\bibfnamefont {D.~L.}\ \bibnamefont
  {Feder}},\ }\href@noop {} {\bibfield  {journal} {\bibinfo  {journal}
  {Physical Review A}\ }\textbf {\bibinfo {volume} {78}},\ \bibinfo {pages}
  {032312} (\bibinfo {year} {2008})}\BibitemShut {NoStop}%
\bibitem [{\citenamefont {Blythe}\ and\ \citenamefont
  {Varcoe}(2006)}]{blythe2006cavity}%
  \BibitemOpen
  \bibfield  {author} {\bibinfo {author} {\bibfnamefont {P.}~\bibnamefont
  {Blythe}}\ and\ \bibinfo {author} {\bibfnamefont {B.}~\bibnamefont
  {Varcoe}},\ }\href@noop {} {\bibfield  {journal} {\bibinfo  {journal} {New
  Journal of Physics}\ }\textbf {\bibinfo {volume} {8}},\ \bibinfo {pages}
  {231} (\bibinfo {year} {2006})}\BibitemShut {NoStop}%
\bibitem [{\citenamefont {Broadbent}\ \emph {et~al.}(2009)\citenamefont
  {Broadbent}, \citenamefont {Fitzsimons},\ and\ \citenamefont
  {Kashefi}}]{broadbent2009}%
  \BibitemOpen
  \bibfield  {author} {\bibinfo {author} {\bibfnamefont {A.}~\bibnamefont
  {Broadbent}}, \bibinfo {author} {\bibfnamefont {J.}~\bibnamefont
  {Fitzsimons}}, \ and\ \bibinfo {author} {\bibfnamefont {E.}~\bibnamefont
  {Kashefi}},\ }in\ \href@noop {} {\emph {\bibinfo {booktitle} {50th Annual
  IEEE Symposium on Foundations of Computer Science, FOCS'09}}}\ (\bibinfo
  {year} {2009})\ pp.\ \bibinfo {pages} {517--526}\BibitemShut {NoStop}%
\bibitem [{\citenamefont {Fitzsimons}\ and\ \citenamefont
  {Kashefi}(2012)}]{fitzsimons2012unconditionally}%
  \BibitemOpen
  \bibfield  {author} {\bibinfo {author} {\bibfnamefont {J.~F.}\ \bibnamefont
  {Fitzsimons}}\ and\ \bibinfo {author} {\bibfnamefont {E.}~\bibnamefont
  {Kashefi}},\ }\href@noop {} {\bibfield  {journal} {\bibinfo  {journal} {arXiv
  preprint arXiv:1203.5217}\ } (\bibinfo {year} {2012})}\BibitemShut {NoStop}%
\bibitem [{\citenamefont {Morimae}\ and\ \citenamefont
  {Fujii}(2013)}]{morimae2013blind}%
  \BibitemOpen
  \bibfield  {author} {\bibinfo {author} {\bibfnamefont {T.}~\bibnamefont
  {Morimae}}\ and\ \bibinfo {author} {\bibfnamefont {K.}~\bibnamefont
  {Fujii}},\ }\href@noop {} {\bibfield  {journal} {\bibinfo  {journal}
  {Physical Review A}\ }\textbf {\bibinfo {volume} {87}},\ \bibinfo {pages}
  {050301} (\bibinfo {year} {2013})}\BibitemShut {NoStop}%
\bibitem [{\citenamefont {Zwerger}\ \emph {et~al.}(2013)\citenamefont
  {Zwerger}, \citenamefont {Briegel},\ and\ \citenamefont
  {D{\"u}r}}]{zwerger2013universal}%
  \BibitemOpen
  \bibfield  {author} {\bibinfo {author} {\bibfnamefont {M.}~\bibnamefont
  {Zwerger}}, \bibinfo {author} {\bibfnamefont {H.}~\bibnamefont {Briegel}}, \
  and\ \bibinfo {author} {\bibfnamefont {W.}~\bibnamefont {D{\"u}r}},\
  }\href@noop {} {\bibfield  {journal} {\bibinfo  {journal} {Physical review
  letters}\ }\textbf {\bibinfo {volume} {110}},\ \bibinfo {pages} {260503}
  (\bibinfo {year} {2013})}\BibitemShut {NoStop}%
\bibitem [{\citenamefont {Zwerger}\ \emph {et~al.}(2014)\citenamefont
  {Zwerger}, \citenamefont {Briegel},\ and\ \citenamefont
  {D{\"u}r}}]{zwerger2014hybrid}%
  \BibitemOpen
  \bibfield  {author} {\bibinfo {author} {\bibfnamefont {M.}~\bibnamefont
  {Zwerger}}, \bibinfo {author} {\bibfnamefont {H.}~\bibnamefont {Briegel}}, \
  and\ \bibinfo {author} {\bibfnamefont {W.}~\bibnamefont {D{\"u}r}},\
  }\href@noop {} {\bibfield  {journal} {\bibinfo  {journal} {Scientific
  reports}\ }\textbf {\bibinfo {volume} {4}} (\bibinfo {year}
  {2014})}\BibitemShut {NoStop}%
\bibitem [{\citenamefont {Raussendorf}\ \emph {et~al.}(2007)\citenamefont
  {Raussendorf}, \citenamefont {Harrington},\ and\ \citenamefont
  {Goyal}}]{raussendorf2007topological}%
  \BibitemOpen
  \bibfield  {author} {\bibinfo {author} {\bibfnamefont {R.}~\bibnamefont
  {Raussendorf}}, \bibinfo {author} {\bibfnamefont {J.}~\bibnamefont
  {Harrington}}, \ and\ \bibinfo {author} {\bibfnamefont {K.}~\bibnamefont
  {Goyal}},\ }\href@noop {} {\bibfield  {journal} {\bibinfo  {journal} {New
  Journal of Physics}\ }\textbf {\bibinfo {volume} {9}},\ \bibinfo {pages}
  {199} (\bibinfo {year} {2007})}\BibitemShut {NoStop}%
\bibitem [{\citenamefont {Kashefi}\ and\ \citenamefont
  {Wallden}(2015)}]{kashefi2015optimised}%
  \BibitemOpen
  \bibfield  {author} {\bibinfo {author} {\bibfnamefont {E.}~\bibnamefont
  {Kashefi}}\ and\ \bibinfo {author} {\bibfnamefont {P.}~\bibnamefont
  {Wallden}},\ }\href@noop {} {\bibfield  {journal} {\bibinfo  {journal} {arXiv
  preprint arXiv:1510.07408}\ } (\bibinfo {year} {2015})}\BibitemShut {NoStop}%
\bibitem [{\citenamefont {Morimae}(2016)}]{morimae2016measurement}%
  \BibitemOpen
  \bibfield  {author} {\bibinfo {author} {\bibfnamefont {T.}~\bibnamefont
  {Morimae}},\ }\href@noop {} {\bibfield  {journal} {\bibinfo  {journal}
  {Physical Review A}\ }\textbf {\bibinfo {volume} {94}},\ \bibinfo {pages}
  {042301} (\bibinfo {year} {2016})}\BibitemShut {NoStop}%
\bibitem [{\citenamefont {Walther}\ \emph {et~al.}(2005)\citenamefont
  {Walther}, \citenamefont {Resch}, \citenamefont {Rudolph}, \citenamefont
  {Schenck}, \citenamefont {Weinfurter}, \citenamefont {Vedral}, \citenamefont
  {Aspelmeyer},\ and\ \citenamefont {Zeilinger}}]{walther2005experimental}%
  \BibitemOpen
  \bibfield  {author} {\bibinfo {author} {\bibfnamefont {P.}~\bibnamefont
  {Walther}}, \bibinfo {author} {\bibfnamefont {K.~J.}\ \bibnamefont {Resch}},
  \bibinfo {author} {\bibfnamefont {T.}~\bibnamefont {Rudolph}}, \bibinfo
  {author} {\bibfnamefont {E.}~\bibnamefont {Schenck}}, \bibinfo {author}
  {\bibfnamefont {H.}~\bibnamefont {Weinfurter}}, \bibinfo {author}
  {\bibfnamefont {V.}~\bibnamefont {Vedral}}, \bibinfo {author} {\bibfnamefont
  {M.}~\bibnamefont {Aspelmeyer}}, \ and\ \bibinfo {author} {\bibfnamefont
  {A.}~\bibnamefont {Zeilinger}},\ }\href@noop {} {\bibfield  {journal}
  {\bibinfo  {journal} {Nature}\ }\textbf {\bibinfo {volume} {434}},\ \bibinfo
  {pages} {169} (\bibinfo {year} {2005})}\BibitemShut {NoStop}%
\bibitem [{\citenamefont {Chen}\ \emph {et~al.}(2007)\citenamefont {Chen},
  \citenamefont {Li}, \citenamefont {Zhang}, \citenamefont {Chen},
  \citenamefont {Goebel}, \citenamefont {Chen}, \citenamefont {Mair},\ and\
  \citenamefont {Pan}}]{chen2007experimental}%
  \BibitemOpen
  \bibfield  {author} {\bibinfo {author} {\bibfnamefont {K.}~\bibnamefont
  {Chen}}, \bibinfo {author} {\bibfnamefont {C.-M.}\ \bibnamefont {Li}},
  \bibinfo {author} {\bibfnamefont {Q.}~\bibnamefont {Zhang}}, \bibinfo
  {author} {\bibfnamefont {Y.-A.}\ \bibnamefont {Chen}}, \bibinfo {author}
  {\bibfnamefont {A.}~\bibnamefont {Goebel}}, \bibinfo {author} {\bibfnamefont
  {S.}~\bibnamefont {Chen}}, \bibinfo {author} {\bibfnamefont {A.}~\bibnamefont
  {Mair}}, \ and\ \bibinfo {author} {\bibfnamefont {J.-W.}\ \bibnamefont
  {Pan}},\ }\href@noop {} {\bibfield  {journal} {\bibinfo  {journal} {Physical
  review letters}\ }\textbf {\bibinfo {volume} {99}},\ \bibinfo {pages}
  {120503} (\bibinfo {year} {2007})}\BibitemShut {NoStop}%
\bibitem [{\citenamefont {Prevedel}\ \emph {et~al.}(2007)\citenamefont
  {Prevedel}, \citenamefont {Walther}, \citenamefont {Tiefenbacher},
  \citenamefont {B{\"o}hi}, \citenamefont {Kaltenbaek}, \citenamefont
  {Jennewein},\ and\ \citenamefont {Zeilinger}}]{prevedel2007high}%
  \BibitemOpen
  \bibfield  {author} {\bibinfo {author} {\bibfnamefont {R.}~\bibnamefont
  {Prevedel}}, \bibinfo {author} {\bibfnamefont {P.}~\bibnamefont {Walther}},
  \bibinfo {author} {\bibfnamefont {F.}~\bibnamefont {Tiefenbacher}}, \bibinfo
  {author} {\bibfnamefont {P.}~\bibnamefont {B{\"o}hi}}, \bibinfo {author}
  {\bibfnamefont {R.}~\bibnamefont {Kaltenbaek}}, \bibinfo {author}
  {\bibfnamefont {T.}~\bibnamefont {Jennewein}}, \ and\ \bibinfo {author}
  {\bibfnamefont {A.}~\bibnamefont {Zeilinger}},\ }\href@noop {} {\bibfield
  {journal} {\bibinfo  {journal} {Nature}\ }\textbf {\bibinfo {volume} {445}},\
  \bibinfo {pages} {65} (\bibinfo {year} {2007})}\BibitemShut {NoStop}%
\bibitem [{\citenamefont {Vallone}\ \emph {et~al.}(2008)\citenamefont
  {Vallone}, \citenamefont {Pomarico}, \citenamefont {De~Martini},\ and\
  \citenamefont {Mataloni}}]{vallone2008active}%
  \BibitemOpen
  \bibfield  {author} {\bibinfo {author} {\bibfnamefont {G.}~\bibnamefont
  {Vallone}}, \bibinfo {author} {\bibfnamefont {E.}~\bibnamefont {Pomarico}},
  \bibinfo {author} {\bibfnamefont {F.}~\bibnamefont {De~Martini}}, \ and\
  \bibinfo {author} {\bibfnamefont {P.}~\bibnamefont {Mataloni}},\ }\href@noop
  {} {\bibfield  {journal} {\bibinfo  {journal} {Physical review letters}\
  }\textbf {\bibinfo {volume} {100}},\ \bibinfo {pages} {160502} (\bibinfo
  {year} {2008})}\BibitemShut {NoStop}%
\bibitem [{\citenamefont {Tokunaga}\ \emph {et~al.}(2008)\citenamefont
  {Tokunaga}, \citenamefont {Kuwashiro}, \citenamefont {Yamamoto},
  \citenamefont {Koashi},\ and\ \citenamefont
  {Imoto}}]{tokunaga2008generation}%
  \BibitemOpen
  \bibfield  {author} {\bibinfo {author} {\bibfnamefont {Y.}~\bibnamefont
  {Tokunaga}}, \bibinfo {author} {\bibfnamefont {S.}~\bibnamefont {Kuwashiro}},
  \bibinfo {author} {\bibfnamefont {T.}~\bibnamefont {Yamamoto}}, \bibinfo
  {author} {\bibfnamefont {M.}~\bibnamefont {Koashi}}, \ and\ \bibinfo {author}
  {\bibfnamefont {N.}~\bibnamefont {Imoto}},\ }\href@noop {} {\bibfield
  {journal} {\bibinfo  {journal} {Physical review letters}\ }\textbf {\bibinfo
  {volume} {100}},\ \bibinfo {pages} {210501} (\bibinfo {year}
  {2008})}\BibitemShut {NoStop}%
\bibitem [{\citenamefont {Yao}\ \emph {et~al.}(2012)\citenamefont {Yao},
  \citenamefont {Wang}, \citenamefont {Chen}, \citenamefont {Gao},
  \citenamefont {Fowler}, \citenamefont {Raussendorf}, \citenamefont {Chen},
  \citenamefont {Liu}, \citenamefont {Lu}, \citenamefont {Deng} \emph
  {et~al.}}]{yao2012experimental}%
  \BibitemOpen
  \bibfield  {author} {\bibinfo {author} {\bibfnamefont {X.-C.}\ \bibnamefont
  {Yao}}, \bibinfo {author} {\bibfnamefont {T.-X.}\ \bibnamefont {Wang}},
  \bibinfo {author} {\bibfnamefont {H.-Z.}\ \bibnamefont {Chen}}, \bibinfo
  {author} {\bibfnamefont {W.-B.}\ \bibnamefont {Gao}}, \bibinfo {author}
  {\bibfnamefont {A.~G.}\ \bibnamefont {Fowler}}, \bibinfo {author}
  {\bibfnamefont {R.}~\bibnamefont {Raussendorf}}, \bibinfo {author}
  {\bibfnamefont {Z.-B.}\ \bibnamefont {Chen}}, \bibinfo {author}
  {\bibfnamefont {N.-L.}\ \bibnamefont {Liu}}, \bibinfo {author} {\bibfnamefont
  {C.-Y.}\ \bibnamefont {Lu}}, \bibinfo {author} {\bibfnamefont {Y.-J.}\
  \bibnamefont {Deng}},  \emph {et~al.},\ }\href@noop {} {\bibfield  {journal}
  {\bibinfo  {journal} {Nature}\ }\textbf {\bibinfo {volume} {482}},\ \bibinfo
  {pages} {489} (\bibinfo {year} {2012})}\BibitemShut {NoStop}%
\bibitem [{\citenamefont {Lanyon}\ \emph {et~al.}(2013)\citenamefont {Lanyon},
  \citenamefont {Jurcevic}, \citenamefont {Zwerger}, \citenamefont {Hempel},
  \citenamefont {Martinez}, \citenamefont {D{\"u}r}, \citenamefont {Briegel},
  \citenamefont {Blatt},\ and\ \citenamefont {Roos}}]{lanyon2013measurement}%
  \BibitemOpen
  \bibfield  {author} {\bibinfo {author} {\bibfnamefont {B.}~\bibnamefont
  {Lanyon}}, \bibinfo {author} {\bibfnamefont {P.}~\bibnamefont {Jurcevic}},
  \bibinfo {author} {\bibfnamefont {M.}~\bibnamefont {Zwerger}}, \bibinfo
  {author} {\bibfnamefont {C.}~\bibnamefont {Hempel}}, \bibinfo {author}
  {\bibfnamefont {E.}~\bibnamefont {Martinez}}, \bibinfo {author}
  {\bibfnamefont {W.}~\bibnamefont {D{\"u}r}}, \bibinfo {author} {\bibfnamefont
  {H.}~\bibnamefont {Briegel}}, \bibinfo {author} {\bibfnamefont
  {R.}~\bibnamefont {Blatt}}, \ and\ \bibinfo {author} {\bibfnamefont
  {C.}~\bibnamefont {Roos}},\ }\href@noop {} {\bibfield  {journal} {\bibinfo
  {journal} {Physical review letters}\ }\textbf {\bibinfo {volume} {111}},\
  \bibinfo {pages} {210501} (\bibinfo {year} {2013})}\BibitemShut {NoStop}%
\bibitem [{\citenamefont {Miwa}\ \emph {et~al.}(2009)\citenamefont {Miwa},
  \citenamefont {Yoshikawa}, \citenamefont {van Loock},\ and\ \citenamefont
  {Furusawa}}]{miwa2009demonstration}%
  \BibitemOpen
  \bibfield  {author} {\bibinfo {author} {\bibfnamefont {Y.}~\bibnamefont
  {Miwa}}, \bibinfo {author} {\bibfnamefont {J.-i.}\ \bibnamefont {Yoshikawa}},
  \bibinfo {author} {\bibfnamefont {P.}~\bibnamefont {van Loock}}, \ and\
  \bibinfo {author} {\bibfnamefont {A.}~\bibnamefont {Furusawa}},\ }\href@noop
  {} {\bibfield  {journal} {\bibinfo  {journal} {Physical Review A}\ }\textbf
  {\bibinfo {volume} {80}},\ \bibinfo {pages} {050303} (\bibinfo {year}
  {2009})}\BibitemShut {NoStop}%
\bibitem [{\citenamefont {Ukai}\ \emph {et~al.}(2011)\citenamefont {Ukai},
  \citenamefont {Yokoyama}, \citenamefont {Yoshikawa}, \citenamefont {van
  Loock},\ and\ \citenamefont {Furusawa}}]{ukai2011demonstration}%
  \BibitemOpen
  \bibfield  {author} {\bibinfo {author} {\bibfnamefont {R.}~\bibnamefont
  {Ukai}}, \bibinfo {author} {\bibfnamefont {S.}~\bibnamefont {Yokoyama}},
  \bibinfo {author} {\bibfnamefont {J.-i.}\ \bibnamefont {Yoshikawa}}, \bibinfo
  {author} {\bibfnamefont {P.}~\bibnamefont {van Loock}}, \ and\ \bibinfo
  {author} {\bibfnamefont {A.}~\bibnamefont {Furusawa}},\ }\href@noop {}
  {\bibfield  {journal} {\bibinfo  {journal} {Physical review letters}\
  }\textbf {\bibinfo {volume} {107}},\ \bibinfo {pages} {250501} (\bibinfo
  {year} {2011})}\BibitemShut {NoStop}%
\bibitem [{\citenamefont {Pooser}\ and\ \citenamefont
  {Jing}(2014)}]{pooser2014continuous}%
  \BibitemOpen
  \bibfield  {author} {\bibinfo {author} {\bibfnamefont {R.}~\bibnamefont
  {Pooser}}\ and\ \bibinfo {author} {\bibfnamefont {J.}~\bibnamefont {Jing}},\
  }\href@noop {} {\bibfield  {journal} {\bibinfo  {journal} {Physical Review
  A}\ }\textbf {\bibinfo {volume} {90}},\ \bibinfo {pages} {043841} (\bibinfo
  {year} {2014})}\BibitemShut {NoStop}%
\bibitem [{\citenamefont {Childs}\ \emph {et~al.}(2005)\citenamefont {Childs},
  \citenamefont {Leung},\ and\ \citenamefont {Nielsen}}]{childs2005unified}%
  \BibitemOpen
  \bibfield  {author} {\bibinfo {author} {\bibfnamefont {A.~M.}\ \bibnamefont
  {Childs}}, \bibinfo {author} {\bibfnamefont {D.~W.}\ \bibnamefont {Leung}}, \
  and\ \bibinfo {author} {\bibfnamefont {M.~A.}\ \bibnamefont {Nielsen}},\
  }\href@noop {} {\bibfield  {journal} {\bibinfo  {journal} {Physical Review
  A}\ }\textbf {\bibinfo {volume} {71}},\ \bibinfo {pages} {032318} (\bibinfo
  {year} {2005})}\BibitemShut {NoStop}%
\bibitem [{\citenamefont {Mantri}\ \emph {et~al.}(2016)\citenamefont {Mantri},
  \citenamefont {Demarie}, \citenamefont {Menicucci},\ and\ \citenamefont
  {Fitzsimons}}]{mantri2016}%
  \BibitemOpen
  \bibfield  {author} {\bibinfo {author} {\bibfnamefont {A.}~\bibnamefont
  {Mantri}}, \bibinfo {author} {\bibfnamefont {T.~F.}\ \bibnamefont {Demarie}},
  \bibinfo {author} {\bibfnamefont {N.~C.}\ \bibnamefont {Menicucci}}, \ and\
  \bibinfo {author} {\bibfnamefont {J.~F.}\ \bibnamefont {Fitzsimons}},\
  }\href@noop {} {\bibfield  {journal} {\bibinfo  {journal} {arXiv preprint
  arXiv:1608.04633}\ } (\bibinfo {year} {2016})}\BibitemShut {NoStop}%
\bibitem [{\citenamefont {Danos}\ \emph {et~al.}(2007)\citenamefont {Danos},
  \citenamefont {Kashefi},\ and\ \citenamefont
  {Panangaden}}]{danos2007measurement}%
  \BibitemOpen
  \bibfield  {author} {\bibinfo {author} {\bibfnamefont {V.}~\bibnamefont
  {Danos}}, \bibinfo {author} {\bibfnamefont {E.}~\bibnamefont {Kashefi}}, \
  and\ \bibinfo {author} {\bibfnamefont {P.}~\bibnamefont {Panangaden}},\
  }\href@noop {} {\bibfield  {journal} {\bibinfo  {journal} {Journal of the ACM
  (JACM)}\ }\textbf {\bibinfo {volume} {54}},\ \bibinfo {pages} {8} (\bibinfo
  {year} {2007})}\BibitemShut {NoStop}%
\bibitem [{\citenamefont {De~Beaudrap}(2008)}]{de2008finding}%
  \BibitemOpen
  \bibfield  {author} {\bibinfo {author} {\bibfnamefont {N.}~\bibnamefont
  {De~Beaudrap}},\ }\href@noop {} {\bibfield  {journal} {\bibinfo  {journal}
  {Physical Review A}\ }\textbf {\bibinfo {volume} {77}},\ \bibinfo {pages}
  {022328} (\bibinfo {year} {2008})}\BibitemShut {NoStop}%
\bibitem [{\citenamefont {Nikahd}\ \emph {et~al.}(2015)\citenamefont {Nikahd},
  \citenamefont {Houshmand}, \citenamefont {Zamani},\ and\ \citenamefont
  {Sedighi}}]{nikahd2015one}%
  \BibitemOpen
  \bibfield  {author} {\bibinfo {author} {\bibfnamefont {E.}~\bibnamefont
  {Nikahd}}, \bibinfo {author} {\bibfnamefont {M.}~\bibnamefont {Houshmand}},
  \bibinfo {author} {\bibfnamefont {M.~S.}\ \bibnamefont {Zamani}}, \ and\
  \bibinfo {author} {\bibfnamefont {M.}~\bibnamefont {Sedighi}},\ }\href@noop
  {} {\bibfield  {journal} {\bibinfo  {journal} {Microprocessors and
  Microsystems}\ }\textbf {\bibinfo {volume} {39}},\ \bibinfo {pages} {210}
  (\bibinfo {year} {2015})}\BibitemShut {NoStop}%
\bibitem [{\citenamefont {Mhalla}\ and\ \citenamefont
  {Perdrix}(2008)}]{mhalla2008finding}%
  \BibitemOpen
  \bibfield  {author} {\bibinfo {author} {\bibfnamefont {M.}~\bibnamefont
  {Mhalla}}\ and\ \bibinfo {author} {\bibfnamefont {S.}~\bibnamefont
  {Perdrix}},\ }in\ \href@noop {} {\emph {\bibinfo {booktitle} {International
  Colloquium on Automata, Languages, and Programming}}}\ (\bibinfo
  {organization} {Springer},\ \bibinfo {year} {2008})\ pp.\ \bibinfo {pages}
  {857--868}\BibitemShut {NoStop}%
\bibitem [{\citenamefont {Browne}\ and\ \citenamefont
  {Briegel}(2006)}]{browne2006one}%
  \BibitemOpen
  \bibfield  {author} {\bibinfo {author} {\bibfnamefont {D.~E.}\ \bibnamefont
  {Browne}}\ and\ \bibinfo {author} {\bibfnamefont {H.~J.}\ \bibnamefont
  {Briegel}},\ }\href@noop {} {\bibfield  {journal} {\bibinfo  {journal} {arXiv
  preprint quant-ph/0603226}\ } (\bibinfo {year} {2006})}\BibitemShut {NoStop}%
\end{thebibliography}%
\end{document}